\documentclass[preprint,prb,amsmath]{revtex4}
\usepackage{graphicx}

\begin{document}


\title{Probing punctual magnetic singularities during magnetization process in FePd films}

\author{Aur\'elien Masseboeuf}
\altaffiliation{New ad. : CNRS, CEMES, F-31055 Toulouse Cedex, France}
\author{Thomas Jourdan}
\altaffiliation{New ad. : CEA, DEN, Service de Recherches de M\'etallurgie Physique, F-91191 Gif-sur-Yvette, France}
\author{Fr\'ed\'eric Lan\c{c}on}
\author{Pascale Bayle-Guillemaud}
\author{Alain Marty}
\affiliation{CEA, INAC, SP2M, F-38054 Grenoble-, France}

\date{\today}

\begin{abstract}
  We report the use of Lorentz microscopy to observe the domain wall
  structure during the magnetization process in FePd thin foils. We have focused
  on the magnetic structure of domain walls of bubble-shaped magnetic domains near saturation. Regions are found along the domain walls where the magnetization abruptly reverses. Multiscale
  magnetic simulations shown that these regions are vertical Bloch lines (VBL) and the different bubble shapes observed are then related to the inner structure of the VBLs. We were thus able to probe the presence of magnetic singularities as small as Bloch points in the inner magnetization of the domain walls.
\end{abstract}
\maketitle

It has been shown that alloys with perpendicular magnetic anisotropy (PMA) 
are good candidates for applications in new recording media with high
density storage capacity or in magneto-logical devices \cite{Weller2000}. Recently such
materials, namely iron palladium (FePd) alloys have been used in spin-valves where they act as the
polarizer and the free layer \cite{Seki2006}. These devices are believed to work through the nucleation of a reversed domain followed by the propagation of a domain
wall \cite{Seki2008}. Numerous studies have advanced the knowledge of the magnetic configuration of FePd alloys by means of MFM imaging \cite{Asenjo1999,Samson1999}, X-ray
scattering \cite{Mulazzi2004}, or numerical
simulations \cite{Toussaint2002}. Moreover it has been shown
recently that it is possible, using Lorentz Transmission Electron
Microscopy (LTEM) to image the magnetic distribution in PMA thin foils at the domain wall scale \cite{Masseboeuf2008}. This method enables quantitative information to be obtained with a spatial resolution below ten nanometers coupled to the opportunity of applying magnetic field during imaging. Here we show that it is also possible to probe the micromagnetic configuration inside the domain wall enabling the detection of defects smaller than the expected spatial resolution.

Along a domain wall one can find some regions where the chirality abruptly switches. The area where the magnetization reverses is called a Bloch line, which can be either horizontal or vertical. An horizontal Bloch line is parallel to the magnetization inside the domain wall while the vertical Bloch line is perpendicular to the magnetization.
Domain walls in PMA materials are Bloch-like walls and the in-plane
magnetization inside the domain wall can thus be oriented in one or the other direction of the wall plane and defines its chirality.
We can thus expect some vertical Bloch lines in FePd delimiting two different chirality of the domain wall.
Vertical Bloch lines were extensively studied in the 80s in
garnets. Experimental \cite{thiaville90,thiaville91,Thiaville2001} and
numerical \cite{Hubert1974,Slonczewski1975,Nakatani1988,Miltat1989}
approaches have helped us to understand these types of magnetic
defects. Observations were possible using magneto-optical
microscopy due to the large width of domain walls in garnets ($\delta
\approx 0.1~\mu\textrm{m}$). This large value has to be compared with
the domain wall width in FePd of around 8 nm \cite{Masseboeuf2009},
well below the resolution of optical methods. The simulation of the magnetic structure in garnets is also much easier than for FePd. Indeed, in garnets the very
high quality factor $Q = 2K/(\mu_0 M_s^2) \approx 8$, where $K$ is the
anisotropy constant and $M_s$ the saturation magnetization, enables
local approximations for the computation of the demagnetizing field \cite{Slonczewski1975}. This assumption is \emph{a priori}
not valid in the case of FePd, which exhibits smaller values of $Q$ in
the order of 1.6.

The aim of this letter is to show that new magnetic modeling coupled
to recent developments in LTEM enables the observation of such small
magnetic defects as VBL in domain walls of less than 10 nm width. We
focus in this work on VBL which are trapped in magnetic bubbles appearing in FePd thin foils near the saturation state.

A thin layer of L1$_0$-FePd (37 nm) has been deposited on a ``soft''
layer of chemically disordered FePd$_2$ layer, grown on a MgO (001)
substrate by Molecular Beam Epitaxy\cite{Beutier2004}. The soft layer
is used to enhance the recording efficiency in perpendicular recording hard
drives (see for example the section 2.4 of \onlinecite{Khizroev2004}). The sample for LTEM has been then prepared using classical
method by mechanical polishing and ion milling. The microscope used is a JEOL 3010 fitted in with a Gatan imaging filter for contrast enhancement by zero-loss filtering \cite{Dooley1997}. The \textit{in-situ}
magnetization is performed with the objective lens while imaging is realized with the objective mini lens traditionally used for low magnification imaging. The field produced by the objective lens has been carefully calibrated by inserting a dedicated sample holder mounted with a Hall probe before the experiment. 

\begin{figure}[!t]
 \includegraphics[width = \columnwidth]{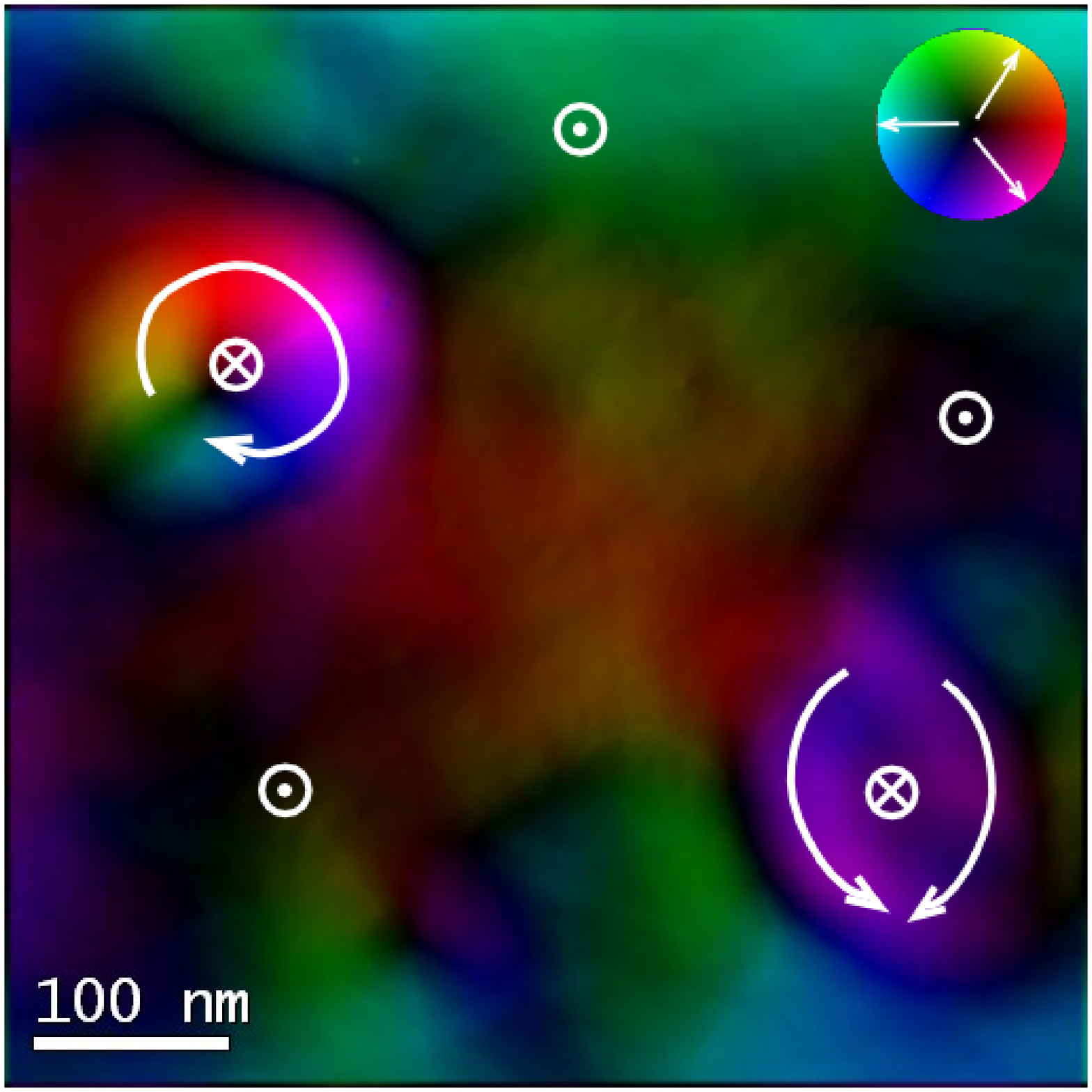}
 \caption{\label{FigTIE} (Colour online) Magnetic induction mapping of
   the FePd thin film using TIE solving at 775 mT. The colour scale
   used here is explained by the colour wheel (colour for the magnetic induction direction and colour intensity for the induction modulus). Arrows are also used to emphasize the magnetic induction. Perpendicular
   induction (\textit{i.e.} magnetization inside the domains) is deduced from the whole magnetization process (saturation state should be up).}
\end{figure}

We measured the half hysteresis loop of the film in Fresnel mode \cite{Chapman1984}. For a field of 775 mT, just before the complete saturation of the magnetic layer, a focal series has been performed. The complete description of this
magnetization process can be found elsewhere\cite{Masseboeuf2008}. Fig.~\ref{FigTIE} shows the focal series reconstruction using the Transport-of-Intensity Equation \cite{Paganin1998}. The magnetic information is originally mixed with an electrostatic contribution \cite{Aharonov1959} which has been removed by considering a constant variation of the thickness of the sample \cite{Masseboeuf2009}. Due to the Lorentz force, the LTEM is sensitive to magnetic induction integrated along the electron beam direction in the TEM. However, assuming that stray fields on both side of the layer are antiparallel, the integrated induction may be considered approximately the same as the integrated magnetization. In the following we will discuss about simulation on integrated magnetization.

In Fig. \ref{FigTIE} we can clearly see two types of magnetic bubbles. On the upper left corner one can see a magnetic bubble where the magnetization swirls continuously along
  the domain wall between the residual magnetic domain (indicated as
  down in Fig. \ref{FigTIE}) and the reversed domain. On the bottom right corner, one finds a magnetic bubble where the magnetization experiences two
  rotations of 180$^{\circ}$ resulting in two ``different'' domain
  walls pointing in similar directions.

At this point of the hysteresis loop a lot of bubbles are
present in the film and both kinds of bubbles can be easily found. The
two switching points observed in the second bubble type are supposed
to be vertical Bloch lines (VBL)\cite{Hubert1998}. It must be noticed
that the geometric deformation observed for the bubble with two VBL (the bubble showing a \emph{lemon} shape) is
fully reproducible.

In order to investigate the internal structure of these lines, we have
performed magnetic simulations on bubbles with and without VBL. Due to the large range of scale needed to model these objects, we developed and used a multiscale efficient method. This method
uses an adaptive mesh refinement technique to achieve both
computational efficiency and numerical accuracy (details
on the method can be found in Ref.~\onlinecite{jourdan08}). This is particularly
useful in the case of bubbles as the inner and outer part of the
bubble can be loosely meshed, whereas the domain wall and the VBL must
be densely meshed\cite{jourdan09}. Moreover, the code we used has the particularity to take into account the atomic structure of the material which is ignored in standard micromagnetic codes. The size of the micromagnetic mesh is then automatically adapted and switch to atomistic mode to keep a good precision when necessary. This ensures that all magnetic configuration is correct as the micromagnetic fundamental assumption of low spatial variations is fulfilled (see \onlinecite{JourdanThesis} for more details and comparisons with traditional code) and at the same time decrease drastically the number of mesh (thus decreasing the calculation time). The following calculation would have cost 8 times more mesh with a traditional micromagnetic parallel code.

In these simulations the saturation magnetization is
$M_s = 10^6~\textrm{A.m}^{-1}$, the anisotropy constant is $K =
10^6~\textrm{J.m}^{-3}$ and the exchange stiffness constant\cite{gehanno97-2} is $A =
7\times 10^{-12}~\textrm{J.m}^{-1}$. With these parameters the
exchange length, $\Lambda = \sqrt{2A/(\mu_0 M_S^2)} =
3.3~\textrm{nm}$. Two different thicknesses have been considered : 15
and 20.7~nm. In Fig.~\ref{VBLsimu} the integrated magnetization along
the thickness obtained from the simulations is shown for a bubble
without VBL (A) and two bubbles with VBL for thicknesses of 15 (B) and 20.7~nm (C)
in a field of 0.25 and 0.3~T respectively.

It can be seen that for a thickness of 15~nm (Fig.~\ref{VBLsimu} B) the bubble is deformed in
agreement with the LTEM observations, whereas for a thickness of
20.7~nm its shape remains circular (Fig.~\ref{VBLsimu} C). 

\begin{figure}[!t]
 \includegraphics[width = \columnwidth]{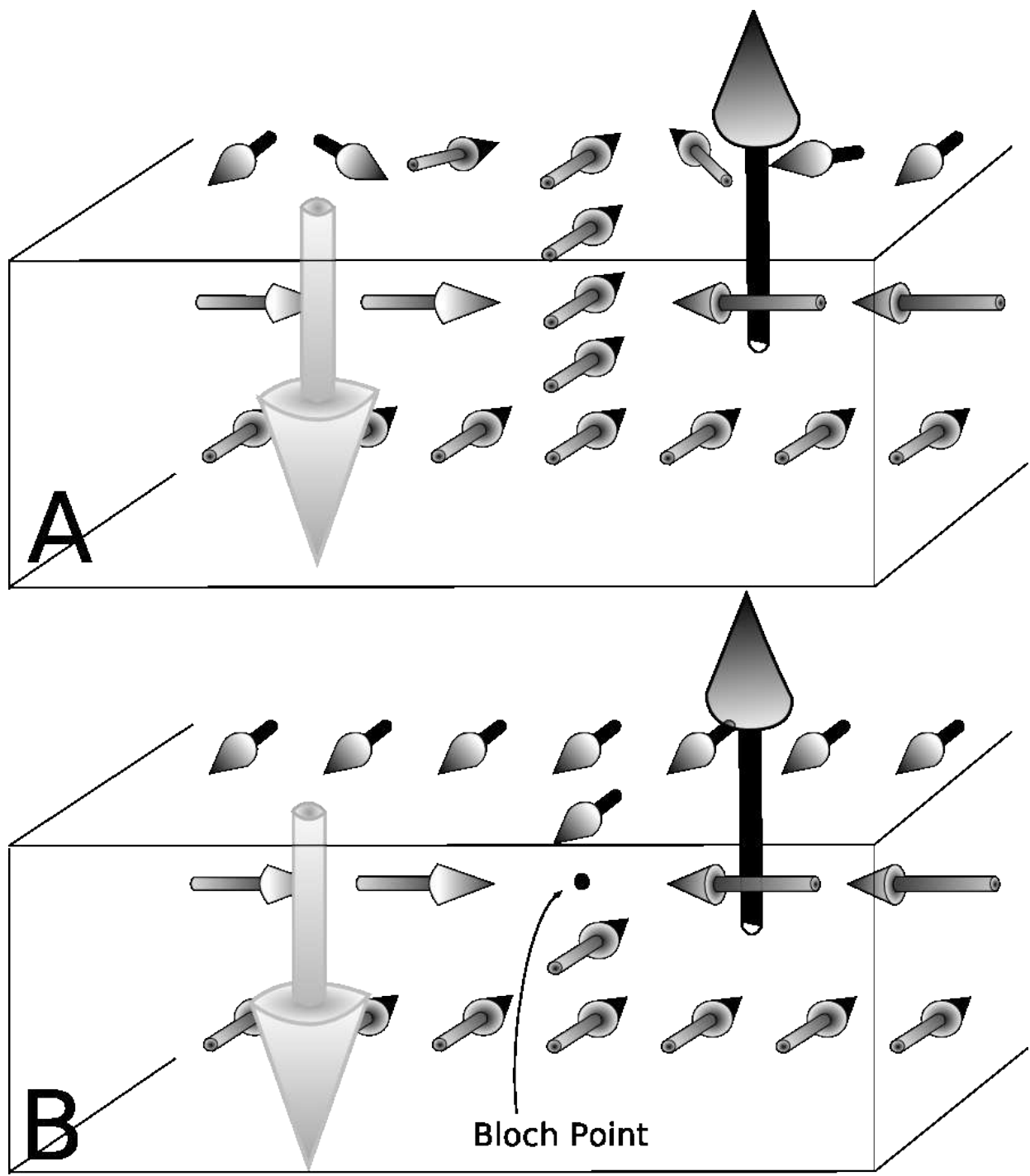}
 \caption{\label{VBLdesc} (A) Vertical Bloch line without Bloch
   point. The upper N\'eel Cap of the Bloch wall is experiencing a
   swirl of 360$^{\circ}$. (B) Vertical Bloch Lines with a
   Bloch point. The N\'eel Caps on each surface remain antiparallel.}
\end{figure}

The modification of the shape can be explained by analysing precisely the structure of the VBL, depending of the thickness. Two kinds of VBL can thus be found. In the case of a small thickness, the magnetization is uniform
along the VBL (Fig.~\ref{VBLdesc} A), whereas it reverses along the
VBL in the second case (large thickness), which leads to a magnetic singularity called a
Bloch point (BP) (Fig.~\ref{VBLdesc} B). The reason for the transition
is a competition between the exchange and demagnetizing energies:
the presence of a Bloch point leads to an increase in the exchange
energy, whereas the demagnetizing energy decreases because the
magnetization in the two segments of the line is aligned along the
stray field generated by the domains.

Such a transition as a function of the thickness, $h$ has been reported by
Hubert to be $h=7.3~\Lambda$ with an analytical model for a straight
domain wall\cite{Hubert1976}. According to our simulations for the
particular geometry considered here, the transition is found between
4.5 and 6.3~$\Lambda$. Given the thickness, $h = 11.2~\Lambda$ of the
films observed by LTEM, the VBL should contain a
Bloch point, which is not consistent with the deformed states
observed. However, the soft layer under the L1$_0$ FePd film changes
the magnetic configuration and alters the respective contributions of
the exchange and demagnetizing terms to the energy. As described in a
previous article\cite{Masseboeuf2007}, the main role of the soft layer
on the domain wall is an enhancement of the size (and as a
consequence, the thickness) of the bottom N\'eel cap of the Bloch walls. This vertical
dissymmetry could thus favours the configuration with no BP by increasing the dipolar energy.

The deformation observed in the absence of a BP gives rise to a
reduction of magnetic charges\cite{jourdan09}: it is analogous to the small
buckling of the magnetization identified in straight domain walls in
garnets\cite{Miltat1989}. In these materials, the buckling reduces the
so-called ``dipolar'' $\pi$ charges which are related to the variation
of the magnetization perpendicular to the domain wall. In the case of
FePd, the lower quality factor, $Q$ reduces the lateral extension of
the VBL, which leads to large ``monopolar'' $\sigma$ charges. A far
larger buckling than could be expected following the studies on
garnets is obtained, beside a reduction of $\pi$ charges, it
also reduces $\sigma$ charges by a compensation of these two
types of charges. It is worthy to note that the magnetization is
oriented in the same direction in both VBL, so that the 360$^{\circ}$-like
domain walls are located on opposite surfaces. To compensate these charges,
a different orientation in the VBL would lead to a ``heart''-shape
bubble, which is not found to be stable in our simulations.

To conclude, in this letter we have highlighted the very high resolution
obtained by combining Lorentz Transmission Electron Microscopy and
multiscale simulations. The resolution we achieved by conventional
electron microscopy enables us to probe magnetic singularities well below the LTEM spatial resolution. Furthermore a main advantage of the multiscale code was its rapidity and its low memory requirements. In that particular case we decrease the number cells thanks to a factor 8 regarding traditional parallel code.  The successful comparison of the two methods shown that it is possible to determine the inner magnetic configuration of a VBL, namely the presence or the absence of Bloch points in them.

\bibliographystyle{apsrev} 
\bibliography{./bubulles}

\newpage
\newpage
\begin{figure}[!t]
 \includegraphics[width = \columnwidth]{VBLsimu}
 \caption{\label{VBLsimu} Magnetic multi-scale simulations of three
   different magnetic bubbles. The magnetization has been integrated
   along the observation direction to correspond to LTEM
   measurements. (A) A magnetic bubble with no VBL. (B)
   A magnetic bubble with two VBL, both VBL contain no Bloch
   point. (C) A magnetic bubble with two VBLs, each VBL
   contains a Bloch point.}
\end{figure}
\newpage
\newpage

\end{document}